# Time Reversal-based Transmissions with Distributed Power Allocation for Two-Tier Networks

Vu Tran-Ha, Quang-Doanh Vu and Een-Kee Hong, *Senior Member, IEEE*
School of Electronic and Information, Kyung Hee University, Yongin, South Korea
E-mail: {tranhavu, vqdoanh, ekhong}@khu.ac.kr

*Abstract*—Radio pollution and power consumption problems lead to innovative development of green heterogeneous networks (HetNet). Time reversal (TR) technique which has been validated from wide- to narrow-band transmissions is evaluated as one of most prominent linear precoders with superior capability of harvesting signal energy. In this paper, we consider a new HetNet model, in which *TR-employed* femtocell is proposed to attain saving power benefits whereas macrocell utilizes the beam-forming algorithm based on zero-forcing principle, over frequency selective channels. In the considered HetNet, the practical case of limited signaling information exchanged via backhaul connections is also taken under advisement. We hence organize a distributed power loading strategy, in which macrocell users are treated with a superior priority compared to femtocell users. By Monte-Carlo simulation, the obtained results show that TR is preferred to zero-forcing in the perspective of beamforming technique for femtocell environments due to very high achievable gain in saving energy, and the validity of power loading strategy is verified over multipath channels.

## I. INTRODUCTION

Heterogeneous network is a prominent approach which satisfies user requirements that traditional cellular network could not be provided [1]-[4]. Femtocell is an effecient solution to provide coverage to deadzones and hotzones in the network. Nowadays, the carrier aggregation technique in long-term evolution-advanced (LTE-A) allows operators to combine up to five carriers as wide as 20MHz each for a maximum bandwidth of 100MHz [2]. And the development of broadband femtocells based LTE-A also brings a cost-efficient provisioning of ubiquitous broad-band services [3]. Otherwise, the growth of energy demand and electromagnetic pollution (e.g. carbon footprint, glass house effect, ect.) yields the development of green cellular networks [4]. In fact, if receivers have a greater diversity performance then BSs can decrease transmit power to reach the green approach whereas the quality-of-service (QoS) is still guaranteed. Equalizers and a more number of antennas are then required, however, high cost and complexity of user equipments are non-preferred.

As an attractive approach to improve the capability of diversity without the increase of the number of antennas, the authors [5]-[7] argued that TR-based transmission is an ideal paradigm for green wireless communications due to its inherent nature of fully harvesting energy from surrounding environment. Through utilizing the time-reversed and conjugated form of channel state information (CSI) to prefilter signals, the TR-based transmissions focus signals tightly in the space and time domains. Thus, the power of all paths is harvested and the signal is compressed in the time and space domains at the receiver.

Although most of previous works focus on TR ultra-wideband system [6]-[10], the feasibility of TR for conventional bandwidth system has been mentioned [5], [11]-[13]. The work [5] discusses the potential of applications of TR in 3GPP LTE-A coordinated multi-point networks. A measured-based investigation of TR, carried out with 10MHz bandwidth centered at 2.14GHz which comparable to the standard of 3G WCDMA systems, revealed the capabilities of temporal and spatial focusing property [12]. The experimental results in [13] also prove that TR properties (i.e. focusing gain, etc.) with narrow bandwidths are very similar to wideband environments. We can conclude that the robustness of TR technique is valid for conventional broadband HetNets.

In this paper, we propose applying TR to two-tier broadband HetNet. In practical, it may not be suitable to apply TR to macrocell due to the high mobility of macro-users [5]. Therefore, we consider in this paper a novel model that macrocell and femtocell base stations implement different beamforming techniques, particularly over frequency selective channels. In flat-fading channels, lots of prior papers for transmit beamforming are carried out [14]-[16], and there have been few works focusing on multi-user multi-tier cellular network with multi-path channels. However, the transmission effects of frequency-selective environments now need to be taken into account due to the expansion of a much wider bandwidth in LTE-A systems [2]. In our model, macrocell base station (MBS) utilizes the proposed beamforming algorithm based on zero-forcing principle, whereas femtocell base station (FBS) employs the conventional TR prefilter, and low complexity users (i.e. equipped one antenna and a single tap diversity combiner) are considered. The main contributions of our paper are that

- Since users are equipped a single tap receiver, we reveal a zero-forcing-based beamforming algorithm for macro-

This work was supported by National Research Foundation of Korea, under Grant Number: 2012K1A3A1A26034927.

cell network to adopt this circumstance. Moreover, to improve energy-efficiency, we propose employing TR technique to femtocell network. The advantages of TR femtocell network are investigated through a comparison between the TR-based and zero-forcing-based beamforming under ITU-R channel model [18].

- We develop a specific distributed power allocation scheme for downlinks in the considered HetNet, under practical assumption of backhaul connections may only support a limited throughput for signalling exchange between two tiers. Our strategy relies on the trade-off between a smaller quantity of demanded CSI and optimal performance.

The rest of the paper is organized as follows: Section II presents the focalization properties of TR technique, the system description is shown in Section II, we construct the problem statements in Section III, the numerical results are provided in Section IV and conclusion is inferred in Section V.

## II. FOCALIZATION PROPERTY OF TR TECHNIQUE IN ITU-R CHANNEL MODEL

In this section, the focalization property is investigated over the femtocell channel model. In multipath channels, we suppose that the maximum length of each channel impulse response (CIR) is $L$. Thus, the CIR between the transmit antenna and the $n$-th user can be presented as

$$h_n(t) = \sum_{l=1}^{L} \alpha_n^{(l)} \delta(t - \tau_n^{(l)}). \quad (1)$$

where $\alpha_n^{(l)}$ and $\tau_n^{(l)}$ are the amplitude and the delay of the $l$-th tap, respectively. We present discrete time form of $h_n(t)$ as

$$\mathbf{h}_n = \begin{bmatrix} \mathbf{h}_n[1] & \mathbf{h}_n[2] & \ldots & \mathbf{h}_n[L] \end{bmatrix}^T, \; \mathbf{h}_n \in \mathbb{C}^{L \times 1} \quad (2)$$

First of all, the transmitter estimates the CIR of channel from intended user by using pilot signals generated from the user equipment. The principle of TR technique lies in the reverse and conjugate of the CIR as a matched filter (pre-filter) at the transmitter side. The pre-filtered signals are back conveyed over the channel which own CIR that used for pre-filtering process. Due to active modulation scheme of TR technique, there are no needs of employing complex equalizers at users. The power of received signal is very tightly focused at one specific time instant and one location at the receiver. Therefore, inter-symbol interference (ISI) and inter-user interference (IUI) are greatly reduced by temporal and spatial focusing respectively. Based on TR scheme, we define $\mathbf{g}_n$ as normalized pre-filterting vector. That is,

$$\mathbf{g}_n[l] = \mathbf{h}_n^*[L+1-l] \Big/ \sqrt{\|\mathbf{h}_n\|^2}. \quad (3)$$

We let $\hat{\mathbf{h}}_{nn'} = \mathbf{g}_n * \mathbf{h}_{n'}$ be equivalent channel, each tap of the equivalent channel might be computed as follows

$$\hat{\mathbf{h}}_{nn'}[k] = \sum_{l=1}^{k} \mathbf{g}_n[l] \mathbf{h}_{n'}[k+1-l] \\ = \sum_{l=1}^{k} \mathbf{h}_n^*[L+1-l] \mathbf{h}_{n'}[k+1-l] \Big/ \sqrt{\|\mathbf{h}_n\|^2}. \quad (4)$$

The simulation is carried out under the system that follows by the description of the indoor channel model in the ITU-R standard [22]. In general, the LTE femtocell channel includes 6 taps and it is illustrated as Fig. 1. In Fig. 2 and Fig. 3, we show the equivalent channels at the intended user and the unintended user respectively. The signal energy is only harvested and compressed in one specific time instant at the desired user, while the signal transmitted for a specific user affects as a noise at other users.

In an ideal case, the peak looks like a Dirac delta function. Nevertheless, the peak becomes wider and some secondary lobes appear in a practical case due to the limitations with respect to bandwidth and hardware equipment [5], [12].

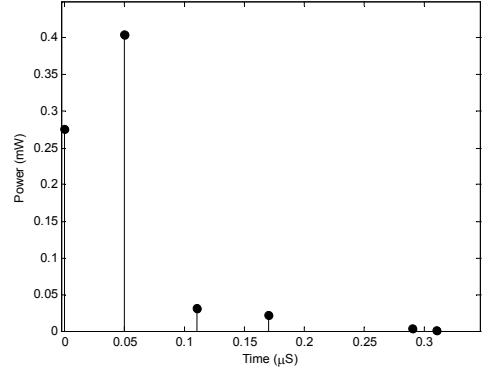

Fig. 1. Channel impulse response.

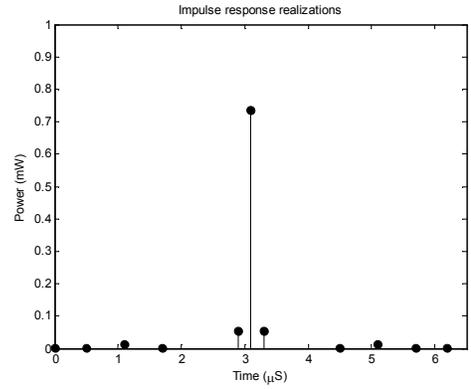

Fig. 2. Equivalent channel at intended user ($n' = n$).

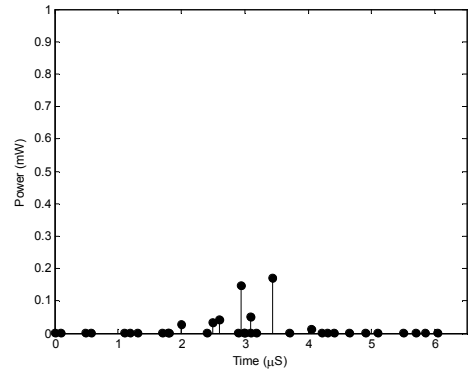

Fig. 3. Equivalent channel at unintended user ($n' \neq n$).

## III. SYSTEM DESCRIPTION

We consider a two-tier HetNet system consisting of one MBS and one FBS where MBS serves $N_0$ macrocell users (MUs) and FBS serves $N_1$ femtocell users (FUs). An example of the considered network is shown in Fig. 4. *For notational convenience, we refer to the MBS as BS 0 and the FBS as BS 1*. The number of antennas equipped at BS $k$ is denoted by $M_k$. We assume that user terminals are low-complexity, i.e. they are equipped with one antenna and a single tap diversity combiner. Let $\mathbf{h}_{ij}^{kr} \in \mathbb{C}^{L \times 1}$ ($0 \le k \le 1$, $0 \le r \le 1$, $0 \le i \le M_k$, $0 \le j \le N_r$) present the CIRs between the $i$-th transmit antenna of the BS $k$ and the $j$-th user of the BS $r$ where $L$ is the number of taps. Moreover, we use superscript $k$ to briefly present superscript $kk$ (i.e $\mathbf{h}_j^0$ is used to denote $\mathbf{h}_j^{00}$).

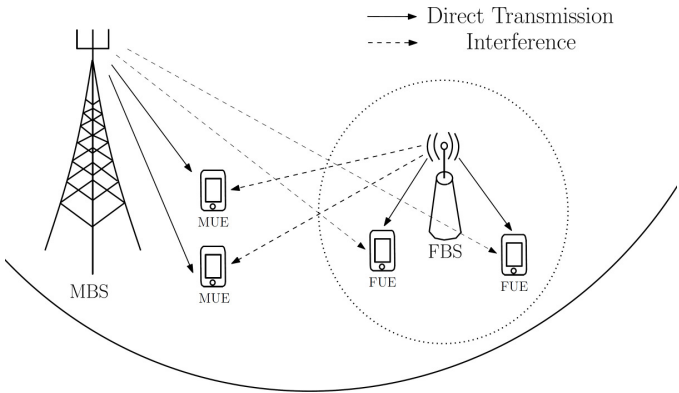

Fig. 4. A two-tier system model including a macrocell and a femtocell.

We recall that, the designed precoder is a combination of transmit power and phase rotation. To deal with frequency selective fading cases, we are interested in decoupling downlink power control and beamforming strategy into distinct processes. Therefore, the transmitted signals at MBS to the $n$-th MU and the FBS to $j$-th FU are given by

$$\mathbf{x}_n^0 = \sqrt{p_n^0}\begin{bmatrix} \mathbf{u}_{0n} & \cdots & \mathbf{u}_{M_0 n} \end{bmatrix}^T s_n^0, \quad (5)$$

$$\mathbf{x}_j^1 = \sqrt{p_j^1}\begin{bmatrix} \mathbf{g}_{0j} & \cdots & \mathbf{g}_{M_1 j} \end{bmatrix}^T s_j^1, \quad (6)$$

respectively, where $s_n^0$ and $s_j^1$ are the transmit signal for the $n$-th MU and the $j$-th FU, respectively. Particularly, $\mathbf{u}_{mn} \in \mathbb{C}^{L \times 1}$ is normalized beamforming vector for the $n$-th MU used at the $m$-th transmit antenna, and $\mathbf{g}_{ij} \in \mathbb{C}^{L \times 1}$ is defined as beamformer for the $j$-th FU at the $i$-th transmit antenna. Additionally, we define $\mathbf{p}^k = \begin{bmatrix} \sqrt{p_1^k} & \sqrt{p_2^k} & \cdots & \sqrt{p_{N_k}^k} \end{bmatrix}^T$ as transmit power vector of BS $k$. With these introduced notations, the received signal at the $n$-th MU can be written as (7). In which, $\mathbf{n}_M$ is Gaussian noise, the first term is the received signal form for the $n$-th MUE, the second term is the co-tier interference in macrocell and the third term is the cross-tier interference. Note that $*$ represents convolution operator. The received signal at the $j$-th FU can be expressed as below

$$\mathbf{y}_n^0 = \sum_{m=1}^{M_0} \sqrt{p_n^0}\mathbf{u}_{mn} * \mathbf{h}_{mn}^0 s_n^0 + \sum_{\substack{n'=1 \\ n' \ne n}}^{N_0} \sum_{m=1}^{M_0} \sqrt{p_{n'}^0}\mathbf{u}_{mn'} * \mathbf{h}_{mn}^0 s_{n'}^0$$
$$+ \sum_{j=1}^{N_1} \sum_{i=1}^{M_1} \sqrt{p_j^1}\mathbf{g}_{ij} * \mathbf{h}_{in}^{10} s_j^1 + \mathbf{n}_M. \quad (7)$$

And the received signal at the $j$-th FU can be expressed as

$$\mathbf{y}_j^1 = \sum_{i=1}^{M_1} \sqrt{p_j^1}\mathbf{g}_{ij} * \mathbf{h}_{ij}^1 s_j^1 + \sum_{\substack{j'=1 \\ j' \ne j}}^{N_1} \sum_{i=1}^{M_1} \sqrt{p_{j'}^1}\mathbf{g}_{ij'} * \mathbf{h}_{ij}^1 s_{j'}^1$$
$$+ \sum_{n=1}^{N_0} \sum_{m=1}^{M_0} \sqrt{p_n^0}\mathbf{u}_{mn} * \mathbf{h}_{mj}^{01} s_n^0 + \mathbf{n}_F, \quad (8)$$

where $\mathbf{n}_F$ is Gaussian noise, the first term is the obtained signal for the $j$-th FU, the second term is the co-tier interference in a same femtocell, the third term is the cross-tier interference from the macrocell.

With the assumed receiver structure, the SINR at the $n$-th MU and the $j$-th FU can be formulated as (9) and (10) shown at the top of next page, respectively. In which, $\alpha$ and $\beta$ denote the position of sampled taps. $P_{(sig)j}^k$, $P_{(isi)j}^k$, $P_{(co)j}^k$, and $P_{(cross)j}^k$ are the desired signal, inter-symbol interference, co-tier and cross-tier interference power, respectively.

## IV. DISTRIBUTED POWER ALLOCATION STRATEGY FOR TR HETNETS

In this section, we are interest in the problem of minimizing total radiated power subject to satisfying the user-specific quality of service (QoS) which is formulated as

$$\begin{aligned} \underset{\{\mathbf{p}^k\}_{k=0}^1}{\text{minimize}} \quad & \sum_{k=0}^1 \|\mathbf{p}^k\|^2 \\ \text{subject to} \quad & \text{SINR}_j^k \ge \gamma_j^k, \quad (\forall k = 0,1; \ 1 \le j \le N_k) \end{aligned} \quad (11)$$

herein $\gamma_j^k$ is preset threshold for user the $j$-th of BS $k$.

With beamformers shown in following subsections, it is not difficult to obtain the optimal solution of (11) due to the fact that it is a convex problem. However, centralized approach requires full CIRs knowledge at central node, and thus the overhead as well as computational burden are heavy to implement in practical, especially when the size of network is large. Moreover, solid backhaul connections must be always available to accommodate BSs with totally substantial information. In cases of the connections damaged, obtaining sufficient CIRs is a great challenge. Comparing to centralized manner, a method which separates the optimization problem into subproblems with a smaller amount of required CSI is much more attractive. As one of our main contributions, a distributed method which yields near optimal performance for considered HetNet is presented as follows.

$$\text{SINR}_n^0\left(\alpha, \mathbf{p}^0, \{\mathbf{u}_{mn}\}_{n=1}^{N_0}\right) = \frac{\underbrace{\left|\left(\sum_{m=1}^{M_0} \sqrt{p_n^0}\mathbf{u}_{mn} * \mathbf{h}_{mn}^0\right)[\alpha]\right|^2}_{P_{(sig)n}^0}}{\underbrace{\sum_{\substack{l=\alpha\\l\neq\alpha}}^{2L-1}\left|\left(\sum_{m=1}^{M_0}\sqrt{p_n^0}\mathbf{u}_{mn}*\mathbf{h}_{mn}^0\right)[l]\right|^2}_{P_{(isi)n}^0} + \underbrace{\sum_{\substack{n'=1\\n'\neq n}}^{N_0}\left\|\sum_{m=1}^{M_0}\sqrt{p_{n'}^0}\mathbf{u}_{mn'}*\mathbf{h}_{mn}^0\right\|^2}_{P_{(co)n}^0} + \underbrace{\sum_{j=1}^{N_1}\left\|\sum_{i=1}^{M_1}\sqrt{p_j^1}\mathbf{g}_{ij}*\mathbf{h}_{in}^{10}\right\|^2}_{P_{(cross)n}^0} + \|\mathbf{n}_M\|^2},$$ (9)

$$\text{SINR}_j^1\left(\beta, \mathbf{p}^1, \{\mathbf{g}_{ij}\}_{i=1,j=1}^{M_1,N_1}, P_{(cross)j}^1\right) = \frac{\underbrace{\left|\left(\sum_{i=1}^{M_1}\sqrt{p_j^1}\mathbf{g}_{ij}*\mathbf{h}_{ij}^1\right)[\beta]\right|^2}_{P_{(sig)j}^1}}{\underbrace{\sum_{\substack{l=1\\l\neq\beta}}^{2L-1}\left|\left(\sum_{i=1}^{M_1}\sqrt{p_j^1}\mathbf{g}_{ij}*\mathbf{h}_{ij}^1\right)[l]\right|^2}_{P_{(isi)j}^1} + \underbrace{\sum_{\substack{j'=1\\j'\neq j}}^{N_1}\left\|\sum_{i=1}^{M_1}\sqrt{p_{j'}^1}\mathbf{g}_{ij'}*\mathbf{h}_{ij}^1\right\|^2}_{P_{(co)j}^1} + \underbrace{\sum_{n=1}^{N_0}\left\|\sum_{m=1}^{M_0}\sqrt{p_n^0}\mathbf{u}_{mn}*\mathbf{h}_{mj}^{01}\right\|^2}_{P_{(cross)j}^1} + \|\mathbf{n}_F\|^2}.$$ (10)

### A. Subproblem 1: Beamforming and power loading problem for TR-applied FBS

In HetNets, MUs usually are the main victims of cross-tier interference, however, they have a strictly greater priority than FUs. As a deduction, the optimization designs must maintain QoS requirements to MUs at all time possible. Following this argument, in our manner, to be specific for MUs's priority, we wish femtocell allocate the radiated power with minimizing interference target for MUs first.

The installation of TR brings capability of mitigating interference to femtocell because of its inherence of location signature-specific property [7], [8]. Note that the reversed-form of CIR is utilized as a matched-filter to prefilter message-bearing signals which is represented as

$$\mathbf{g}_{ij}[l] = \mathbf{h}_{ij}^{*1}[L+1-l] \Big/ \sqrt{\sum_{i=1}^{M_1}\|\mathbf{h}_{ij}^1\|^2}.$$ (12)

In conventional TR, $\mathbf{g}_{ij}$ is appraised as FBS beamformer and the receiver thus selects the central tap (i.e. $\beta = L$) to take a sample. For initiation, femtocell starts with optimizing power allocation with the given tolerable level of cross-interference per each user $P_{(tol)j}^{01}$. The downlink power control problem of the FBS is formulated as below

$$\underset{\{\mathbf{p}^1\}}{\text{minimize}} \quad \sum_{j=1}^{N_1}\left(\sum_{n=1}^{N_0}\left\|\sum_{i=1}^{M_1}\mathbf{g}_{ij}*\mathbf{h}_{in}^{10}\right\|^2 p_j^1\right)$$ (13)

subject to $\text{SINR}_j^1\left(L,\mathbf{p}^1,\{\mathbf{g}_{ij}\}_{i=1,j=1}^{M_1,N_1}, P_{(tol)j}^{01}\right) \geq \gamma_j^1$,

whose optimal solution can be solved by convex optimization techniques with CVX solver [17]. This minimization problem precisely requires the CSI knowledge of femtocell only. After the optimization problem (13) is tackled, the optimal beamformer $\left(\mathbf{p}^1, \{\mathbf{g}_{ij}\}_{i=1,j=1}^{M_1,N_1}\right)$ and $\{\mathbf{h}_{in}^{10}\}_{j=1,n=1}^{N_1,N_0}$ are sent via backhaul to the macrocell.

### B. Subproblem 2: Beamformer design with single tap transmit diversity and power allocation problem for MBS

The optimal beamforming vectors for MBSs are first chosen following *zero-forcing* technique, and the power allocation is hence carried out to maximize corresponding SINR.

In fact, the total received signal is summation of signals from transmit antennas propagated via distinct paths to the intended user. Since assumed users have low complexity and a single tap diversity combiner is employed in MU and FU receivers, therefore, it only takes one sample at a particular tap.

In our algorithm, we specifically observe the *l-th tap* and *other taps* as the *desired signal* and *ISI terms*, respectively, and the corresponding beamformer must be designed following zero-forcing scheme. Thereupon, with equivalent channels including $2L-1$ taps, we obtain $2L-1$ relevant beamformers. Finally, we find out the best among of these candidates. To tackle this problem, we develop *Algorithm 1* to provide the solution.

We start with beamformer design based on the well-known zero-forcing technique over macrocell environments. We let

$$\overline{\mathbf{h}}_{ln} = \begin{bmatrix} \mathbf{h}_{1n}^0[l] & \mathbf{h}_{2n}^0[l] & \cdots & \mathbf{h}_{M_0n}^0[l] \end{bmatrix} \in \mathbb{C}^{1\times M_0}.$$ (14)

Thus, we define $\overline{\mathbf{H}}_n \in \mathbb{C}^{(2L-1)\times M_0 L}$ as

$$\overline{\mathbf{H}}_n = \begin{bmatrix} \overline{\mathbf{h}}_{1n} & \mathbf{0} & & \mathbf{0} \\ \overline{\mathbf{h}}_{2n} & \overline{\mathbf{h}}_{1n} & & \\ \vdots & \overline{\mathbf{h}}_{2n} & & \\ \overline{\mathbf{h}}_{Ln} & \vdots & \ddots & \overline{\mathbf{h}}_{1n} \\ & \overline{\mathbf{h}}_{Ln} & & \overline{\mathbf{h}}_{2n} \\ & & \ddots & \vdots \\ \mathbf{0} & & & \overline{\mathbf{h}}_{Ln} \end{bmatrix}.$$ (15)

For each sampled tap $\alpha$-th, we can arrange corresponding MBS beamformer for *n*-th MU as a $ML\times 1$ vector as follows

$$\overline{\mathbf{w}}_{n,\alpha} = \begin{bmatrix} \overline{\mathbf{u}}_{1n,\alpha}[1] & \cdots & \overline{\mathbf{u}}_{1n,\alpha}[L] & \cdots & \overline{\mathbf{u}}_{M_0n,\alpha}[1] & \cdots & \overline{\mathbf{u}}_{M_0n,\alpha}[L] \end{bmatrix}^T.$$ (16)

Relied on zero-forcing principle, the closed-form derivation of $\bar{\mathbf{w}}_{n,\alpha}$ can be obtained as

$$\bar{\mathbf{w}}_{n,\alpha} = c_{n,\alpha} \left( \begin{bmatrix} \bar{\mathbf{H}}_1^T & \bar{\mathbf{H}}_2^T & \ldots & \bar{\mathbf{H}}_{N_0}^T \end{bmatrix}^T \right)^\dagger \mathbf{z}_{n,\alpha}, \quad (17)$$

where $c_{n,\alpha}$ is a constant to make the norm of $\bar{\mathbf{w}}_n$ is equal to 1, $\mathbf{z}_{n,\alpha} = \begin{bmatrix} \mathbf{0}^T & \ldots & \mathbf{0}^T & \mathbf{s}_\alpha & \mathbf{0}^T & \ldots & \mathbf{0}^T \end{bmatrix}^T$, in which $\mathbf{s}_\alpha = \begin{bmatrix} 0 & \ldots & 0 & 1 & 0 & \ldots & 0 \end{bmatrix}^T \in \mathbb{C}^{(2L-1)\times 1}$ (the 1 is located at $\alpha$-th index), and $(\bullet)^\dagger$ denotes Moore-Penrose pseudo-inverse operator. And from (12), we may infer $\bar{\mathbf{u}}_{mn,\alpha}$. To pre-assess performance, we define a factor $\Gamma_{n,\alpha}$ as

$$\Gamma_{n,\alpha} = \frac{\left| \left( \sum_{m=1}^{M_0} \bar{\mathbf{u}}_{mn,\alpha} * \mathbf{h}_{mn}^0 \right)[\alpha] \right|^2}{\sum_{l\neq\alpha}^{2L-1} \left| \left( \sum_{m=1}^{M_0} \bar{\mathbf{u}}_{mn,\alpha} * \mathbf{h}_{mn}^0 \right)[l] \right|^2 + \sum_{n'\neq n}^{N_0} \sum_{l=1}^{2L-1} \left| \left( \sum_{m=1}^{M_0} \bar{\mathbf{u}}_{mn,\alpha} * \mathbf{h}_{mn'}^0 \right)[l] \right|^2 + 1}. \quad (18)$$

Finally, the procedure finding $\mathbf{u}_{mn}$ is described in *Algorithm 1* as follows

---
**Algorithm 1**: Algorithm to solve $\mathbf{u}_{mn}$

(i). Set $\bar{\alpha} = 1$.
(ii). **Loop**
  1. Compute $\bar{\mathbf{w}}_{n,\bar{\alpha}}$ by (17).
  2. Determine corresponding $\bar{\mathbf{u}}_{mn,\bar{\alpha}}$ by (16), then calculate $\Gamma_{n,\bar{\alpha}}$ by (18).
  3. Update $\bar{\alpha} \leftarrow \bar{\alpha} + 1$.
  **Until** $\bar{\alpha} = 2L - 1$.
(iii). Find $\alpha$ with
  $\alpha = \{\bar{\alpha} \mid \Gamma_{n,\bar{\alpha}} = \max\{\Gamma_{n,1}, \ldots, \Gamma_{n,2L-1}\}\}$.
(vi). Let $\mathbf{w}_n = \bar{\mathbf{w}}_{n,\alpha}$.
(vii). Chosen beamformer $\mathbf{u}_{mn}$ can be inferred following $\mathbf{w}_n$ and (16).

---

After obtaining $\left( \mathbf{p}^1, \{\mathbf{g}_{ij}\}_{i=1,j=1}^{M_1,N_1} \right)$ and $\{\mathbf{h}_{in}^{10}\}_{j=1,n=1}^{N_1,N_0}$, we determine power allocation strategy by computing optimal value of $\mathbf{p}^0$. The convex optimization problem with the SINR and interference constraints is formulated as

$$\begin{aligned} \underset{\{\mathbf{p}^0\}}{\text{minimize}} \quad & \|\mathbf{p}^0\|^2 \\ \text{subject to} \quad & \text{SINR}_n^0 \left( \alpha, \mathbf{p}^0, \{\mathbf{u}_{mn}\}_{m=1,n=1}^{M_0,N_0}, \{\mathbf{h}_{in}^{10}\}_{j=1,n=1}^{N_1,N_0} \right) \geq \gamma_n^0, \\ & \left\| \sum_{m=1}^{M_0} \mathbf{u}_{mn} * \mathbf{h}_{mj}^{01} \sqrt{p_n^0} \right\|^2 \leq P_{(tol)j}^{01}, \end{aligned} \quad (19)$$

and this problem can be tackled by CVX solver [19].

In this section, we have introduced the beamformer designs and the energy loading manners for the macrocell as well as femtocell in which MUs are treated with a distinguished priority. The most fascination of the suboptimal approach lies in the trade-off between a much lower amount of demanded CSI and near optimal performance. Additionally, femtocell itself might flexibly adopt changes of the network as well as the computational burden is appreciably scaled down.

## V. NUMERICAL RESULTS AND DISCUSSIONS

We analyze the transmission power to demonstrate the effectiveness of our proposals by Monte-Carlo simulation. For the HetNet including one MBS and one FBS, the radius of the MBS and FBS are 200m and 10m respectively. FBS are uniformly located in a circle of 100m far from MBS. MUs and FUs are also uniformly allotted in the served areas of MBS and FBSs, respectively. The simulation is carried out with 1000 random locations of users in the considered HetNet environment. We set $\|\mathbf{n}_M\|^2 = \|\mathbf{n}_F\|^2 = 1$ and other parameters are listed in Table 1.

TABLE I. SIMULATION PARAMETER.

| Number of taps, $L$ | 6 | Number of users at FBS | 2 |
|---|---|---|---|
| Number of antennas at MBS | 4 | Pathloss exponent of outdoor, indoor, outdoor to indoor link | 4, 3, 3.5 |
| Number of antennas at FBS | 4 | $P_{(tol)j}^{01}$ | $-7$dBm |
| Number of users at MBS | 2 | $P_{(tol)n}^{10}$ | $-7$dBm |

Taking a remark that, for HetNet channels, we implement the ITU-R channel model [18] that is relevant to cellular systems operating with 20MHz bandwidth (i.e. vehicular and indoor channels for macrocell and femtocell environments, respectively).

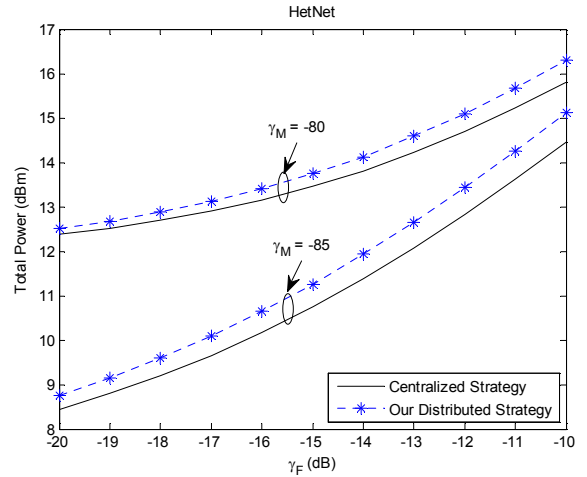

Fig. 5. Performance of power allocation strategies.

For conveniences, we let $\{\gamma_j^1\}_{j=1}^{N_1} = \gamma_F$ and $\{\gamma_n^0\}_{n=1}^{N_0} = \gamma_M$. Fig. 5 shows the pricing of the distributed design for the HetNet with TR femtocell utilized. In Fig. 2, there has been a compromise between two approaches. From observations, we evaluate that the power gap at $\gamma_F = -10$dB is roughly 0.5dB and 0.7dB in cases of $\gamma_M = -80$dB and $-85$dB respectively. The validation of our strategy is then testified

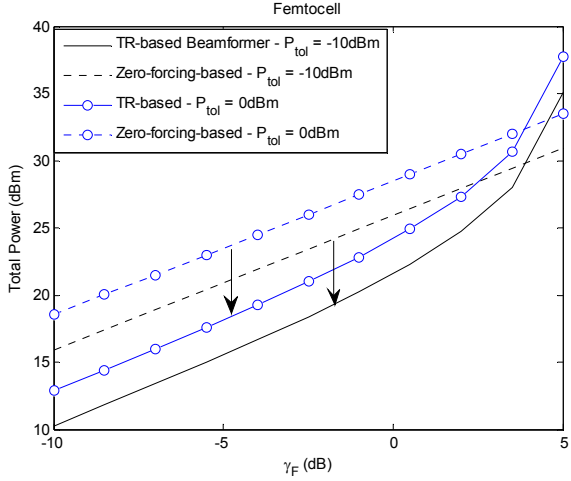

Fig. 6. A comparison between TR and zero-forcing.

In Fig. 6, we compare the effectiveness of TR-based beamformer with that of zero-forcing-based beamforming design (*Algorithm 1*). Under this circumstance, we establish the optimization problem as follows

$$\begin{aligned} \underset{\{\mathbf{p}^1\}}{\text{minimize}} \quad & \|\mathbf{p}^1\|^2 \\ \text{subject to} \quad & \text{SINR}_j^1\left(\mathbf{p}^1, \{\mathbf{z}_{ij}\}_{i=1,j=1}^{M_1,N_1}, P_{(tol)j}^{01}\right) \geq \gamma_j^1, \end{aligned} \quad (20)$$

where $\mathbf{z}_{ij}$ is the beamforming vector, in which, it can follow either TR or zero-forcing design. For channels, the ITU-R indoor model is still utilized, however, we fix the distance between FBS and FUs (i.e. 7 metre) for the simplicity in simulation.

It can be seen that with TR beamformer we can save a maximum gain of 5.5dBm in radiated power compared to zero-forcing-based algorithm at a SNR region from $-10$dB to 4dB. On the other hand, the beamforming scheme of *Algorithm 1* outperforms that of TR at the entire range of SNR. We can conclude that for working point which induces a low emitted power range, TR beamforming outperforms zero-forcing and converse holds for high radiated energy region. Specifically, in practice, the radiated power of FBS is limited up to 20dBm because of the effects of co- and cross-tier interference [19]. Therefore, TR beamforming technique is more desirable than zero-forcing in femtocell environments.

## VI. CONCLUSIONS

Throughout this paper, we consider (i) the application of TR technique for femtocell networks, (ii) the beamforming algorithm based on zero-forcing for macrocell network, as well as (iii) the novel distributed power allocation supporting the case of limited signaling exchange in HetNets over frequency-selective channels. Accordingly, we tackle the optimization problems of downlink power control for both macrocell and femtocell under frequency selective effects. For the proposed allocation scheme, a small pricing of transmission performance reveals our scheme in a promising approach. Moreover, the simulation results also indicate that the TR outperforms zero-forcing beamforming in femtocell environments with the low complex users.